\documentclass[preprint,twocolumn]{aastex631}

\newcommand\chaobj{Cha 1107-7626}

\newcommand\microns{$\mu$m}

\newcommand\methane{CH$_\mathrm{4}$}
\newcommand\ethylene{C$_\mathrm{2}$H$_\mathrm{4}$}

\shorttitle{Hydrocarbons in Brown Dwarf Disk}
\shortauthors{Flagg et al.}

\graphicspath{{./}{figures/}}

\begin{document}

\title{Detection of Hydrocarbons in the Disk around an Actively-Accreting Planetary-Mass Object}


\correspondingauthor{Laura Flagg}
\email{laura.s.flagg@gmail.com}

\author[0000-0001-6362-0571]{Laura Flagg}
\affiliation{Department of Physics \& Astronomy, Johns Hopkins University,  Baltimore, MD, 21218, USA}

\author[0000-0001-8993-50531]{Aleks Scholz}
\affiliation{SUPA, School of Physics \& Astronomy, University of St Andrews, North Haugh, St Andrews, KY169SS, United Kingdom}

\author[0000-0002-4945-9483]{V. Almendros-Abad}
\affiliation{Istituto Nazionale di Astrofisica (INAF) - Osservatorio Astronomico di Palermo, Piazza del Parlamento 1, 90134, Palermo, Italy}

\author[0000-0001-5349-6853]{Ray Jayawardhana}
\affiliation{Department of Physics \& Astronomy, Johns Hopkins University,  Baltimore, MD, 21218, USA}

\author{Belinda Damian}
\affiliation{SUPA, School of Physics \& Astronomy, University of St Andrews, North Haugh, St Andrews, KY169SS, United Kingdom}

\author[0000-0002-7989-2595]{Koraljka Mu\v{z}i\'c}
\affiliation{Instituto de Astrof\'{i}sica e Ci\^{e}ncias do Espaço, Faculdade de Ci\^{e}ncias, Universidade de Lisboa, Ed. C8, Campo Grande, 1749-016 Lisbon, Portugal}

\author{Antonella Natta}
\affiliation{School of Cosmic Physics, Dublin Institute for Advanced Studies, 31 Fitzwilliam Place, Dublin 2, Ireland}

\author[0000-0001-8764-1780]{Paola Pinilla}
\affiliation{Mullard Space Science Laboratory, University College London, Holmbury St Mary, Dorking, London, UK}

\author{Leonardo Testi}
\affiliation{Dipartimento di Fisica e Astronomia, Università di Bologna, Via Gobetti 93/2, 40122, Bologna, Italy}

\begin{abstract}

We present the 0.6--12-micron spectrum of Cha\,1107-7626, a 6-10 Jupiter-mass free-floating object in the $\sim$2\,Myr-old Chamaeleon-I star-forming region, from observations with the NIRSpec and MIRI instruments onboard the James Webb Space Telescope.  We confirm that Cha\,1107-7626 is one of the lowest-mass objects known to harbor a dusty disk with infrared excess emission at wavelengths beyond 4 microns. Our NIRSpec data, and prior ground-based observations, provide strong evidence for ongoing accretion through Hydrogen recombination lines. In the mid-infrared spectrum, we detect unambiguously emission lines caused by methane (\methane) and ethylene (\ethylene) in its circum-substellar disk. Our findings mean that Cha 1107-7626 is by far the lowest-mass object with hydrocarbons observed in its disk. The spectrum of the disk looks remarkably similar to that of ISO-ChaI 147, a very low mass star with a carbon-rich disk that is 10 to 20 times more massive than Cha\,1107-7626. The hydrocarbon lines can be accounted for with a model assuming gas temperatures of a few hundred Kelvin in the inner disk. The obvious similarities between the spectra of a low-mass star and a planetary-mass object indicate that the conditions in the inner disks can be similar across a wide range of central object masses. 
\end{abstract}


\section{Introduction}

Planets are born in disks around young stars. These protoplanetary disks have been identified for central objects across the entire range of stellar and substellar masses, including some that have masses comparable to giant planets \citep{Natta2001ExploringBrownDwarf, jayawardhanaDiskCensusYoung2003, LuhmanDiscoveryYoungSubstellar2006}. Until very recently, most of the work on substellar disks has been focused on the dust component, by necessity, since its thermal emission is straightforward to detect as infrared excess. The suite of instruments onboard the James Webb Space Telescope (JWST) \citep{Rigby2023SciencePerformanceJWST} now provide us with an excellent opportunity to study the gas in the inner disks of young stars and brown dwarfs. 

Here we present the first detection of mid-infrared emission lines originating in the inner disk around a young free-floating planetary-mass object. Our target, \chaobj, has an estimated mass of 6-10 $M_\mathrm{Jup}$, and is one of the lowest mass objects known to have infrared excess emission from a circum-substellar disk as well as evidence of gas accretion \citep{Luhman2008DiskPopulationChamaeleon, almendros-abadYouthAnalysisInfrared2022}. In the JWST/MIRI spectrum, we find strong emission from the hydrocarbons methane and ethylene, molecules that have previously been identified in disks around stars down to masses of $\sim$0.1 M$_\sun$ \citep{taboneRichHydrocarbonChemistry2023,arabhaviAbundantHydrocarbonsDisk2024}. Thus for the first time we have the opportunity to probe the physical conditions of the warm molecular gas in a disk around an object with a $>$10 times lower mass and a $>$100 times lower luminosity than known so far.

\begin{figure*}[ht!]
\centering
\includegraphics[width=1.0\textwidth]{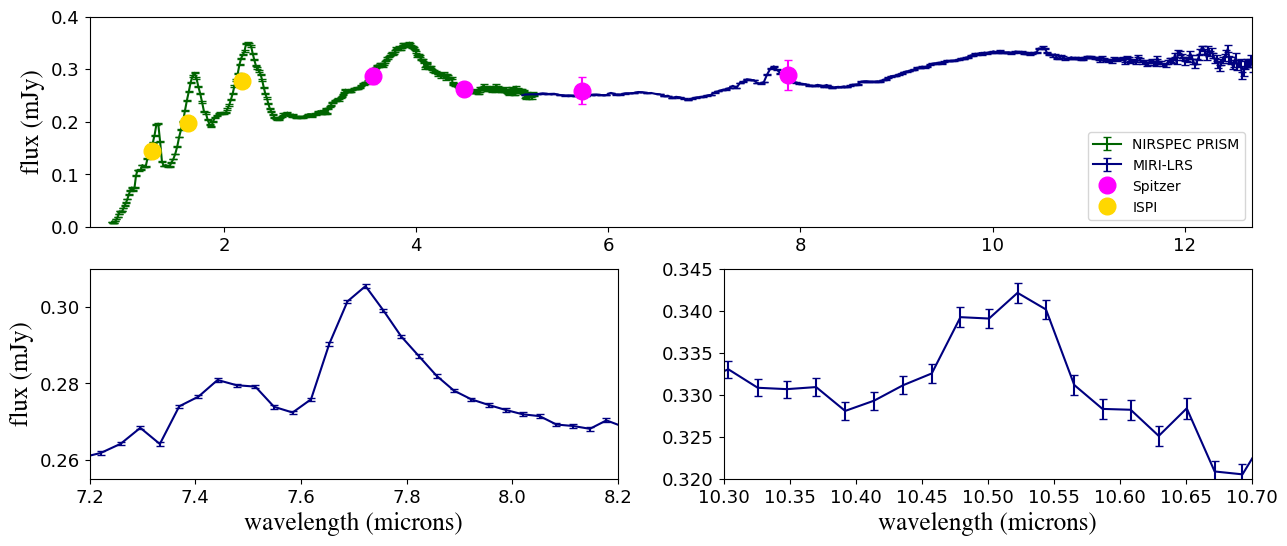}
\caption{(top) The JWST spectrum of \chaobj\ taken with NIRSpec-PRISM (green) and MIRI-LRS (blue). Overplotted are the ISPI and Spitzer photometric points from \citet{Luhman2008DiskPopulationChamaeleon}. (bottom) The spectral data plotted above, zoomed-in on two spectral features in the mid-IR, which we identify as methane and ethylene (see Section \ref{sec:disk}). \label{fig:phot}}
\end{figure*}

\section{Background on Cha 1107-7626}

Cha J11070768-7626326 (hereafter \chaobj) was identified by \citet{Luhman2008DiskPopulationChamaeleon}. In the discovery paper, the authors show evidence of infrared excess from Spitzer photometry and of accretion from H$\alpha$ emission in an optical spectrum. A recent compilation of young brown dwarf spectra by \citet{almendros-abadYouthAnalysisInfrared2022} included a near-infrared VLT/SINFONI spectrum of our target, with a resolution of 1500-2000.

\citet{Luhman2008DiskPopulationChamaeleon} classified the object as L0$\pm1$, with T$_\mathrm{eff}$ of 2200 K and a bolometric luminosity of 3.3 $\times$ 10$^{-4}$ L$_\sun$.  Based on those figures, they estimated a mass of 0.004-0.01 M$_\sun$. In this paper we confirm and refine the estimates for the substellar parameters. 

The object is located in the northern part of the 1-3\,Myr-old Chamaeleon-I star forming region. According to the most recent estimates based on Gaia DR2, this region is at $\sim 190$\,pc \citep{Almendros-Abad2024EvolutionRelationMass}, but earlier studies typically adopted a distance closer by 20-30\,pc \citep{Roccatagliata2018DoublePopulationChamaeleon}. 

\section{Observation and data}
\label{sec:obs}

We observed \chaobj\ with JWST on 2024-08-21 using NIRSpec-PRISM  and MIRI-LRS \dataset[doi:10.17909/jnt4-qg74]{http://dx.doi.org/10.17909/jnt4-qg74}. The instruments achieve a resolution of $R\sim 100$. These observations were part of the GO cycle 3 program 4583 (PI: A. Scholz). The JWST spectra were reduced with pipeline version 1.15.1 \added{\citep{Bushouse2024JWSTCalibrationPipelinev1p15p1}} using the default settings with the exception that we chose to use \textit{pixel\_replace}. The NIRSpec and MIRI spectra match each other at a wavelength of 5$\,\mu m$; they also match the available photometry within the errorbars, indicating that the calibration is robust. The errorbars are directly from the pipeline, and thus only include random noise and not the several percent uncertainty in absolute flux calibration. We removed outliers that differed by more than 5\% from both neighboring points and replaced those values with the average of the two neighboring pixels; this only affected two data points at wavelengths short of 12 \microns.  The full dataset is shown in Figure \ref{fig:phot}. In the near-infrared, the spectrum shows the molecular absorption bands typical for young late-type objects. In the mid-infrared, the spectrum is flat within $\pm 20\%$ with notable emission features at 7.7 and 10.5$\,\mu m$ \added{at S/N of 90 and 13.8 above the continuum, respectively}.  We show plots zoomed-in on those features in the bottom half of Figure \ref{fig:phot}. We note that the spectrum is affected by excessive noise beyond $\lambda=12\,\mu m$. We do not use this part of the spectrum in the following. As seen in Figure \ref{fig:phot}, there may be another spectral feature between 12-12.5$\,\mu m$ (see Sec. 5.1). \added{An emission spike at 12.8$\,\mu m$ is the result of a single pixel with excess flux in only one of the two dithers;} we thus believe it is caused by cosmic ray contamination.

\section{The central object}

\subsection{Basic parameters}
\label{sec:para}

\begin{figure}[ht!]
\centering
\includegraphics[width=0.48\textwidth]{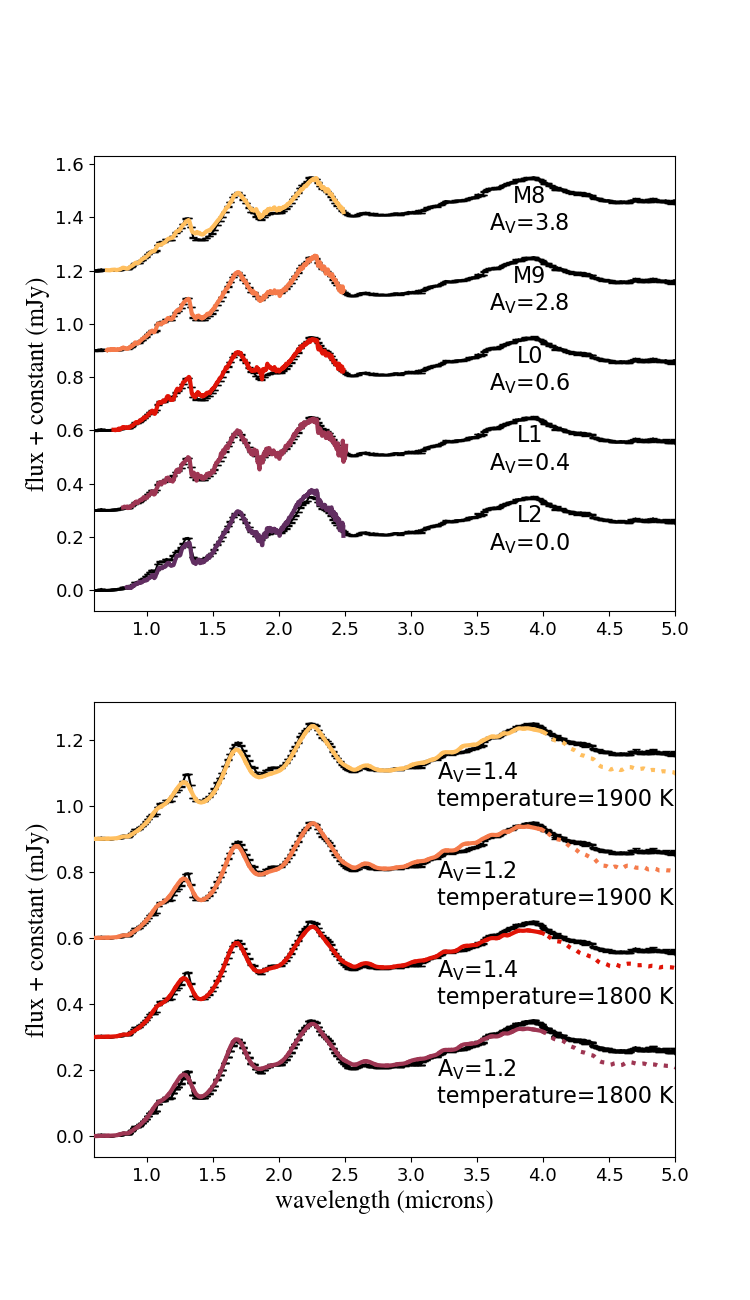}
\caption{(top) Empirical spectral templates in comparison with the NIRSpec data. \added{Templates between M9 and L1 match our data best.  At earlier SpTs ($\leq$M8.5), the slope of the H-band between 1.5 and 1.8 $\mu m$ is too flat, indicating that the source is later type. Whereas at later SpTs ($\geq$L2), the templates overestimate the flux in the J-band, presenting an overall worse fit than in the M9-L1 range.} (bottom) BT-Settl model spectra matched to the NIRSpec data. Models with T$_\mathrm{eff}$ between 1800 and 2000 K provide a good fit to the data. \label{fig:photosphere_fits}}
\end{figure}

We re-measured the spectral type (SpT) using the NIRSpec data, by comparing with spectral templates of young objects. We followed a procedure very similar to that of \citet{Langeveld2024JWSTNIRISSDeep}. In short,  we fit a set of templates to our spectrum, with SpT and extinction as free parameters, normalising the template at a wavelength of 1.66$\,\mu m$. Here we use the extinction law by \citet{Wang2019OpticalMidinfraredExtinction}. We find that templates in the M9-L1 range provide a good fit (see Figure \ref{fig:photosphere_fits}, top), confirming the previously determined SpT. The extinction for these types would be $A_V = 0.4-2.8 $\,mag.

\begin{deluxetable}{lc}
\tablecaption{The physical parameters measured for the \chaobj\  system.\label{tab:parameters}}
\tablehead{%
    \colhead{Parameter} & \colhead{Value} }
\startdata
Mass & 6-10 M$_\mathrm{Jup}$ \\
Effective Temperature & $1900\pm100$ K \\
Spectral Type & L0$\pm1$ \\
V-band Extinction  & $\sim$1\,mag  \\
Log Accretion Luminosity &   $-4.6\pm0.3$ $L_\sun$ \\
Log Accretion Rate Pa$\beta$ & $-9.7 \log{(M_\sun yr^{-1})}$ \\
Log Accretion Rate H$\alpha$ & $-10.6 \log{(M_\sun yr^{-1})}$ \\
\enddata
\end{deluxetable}

We also compare BT-Settl models \citep{AllardAtmospheresVeryLowmass2012} to the NIRSpec spectrum between 0.7 and 4.0\,$\mu m$. A model with T$_\mathrm{eff}=1900\pm 100$\,K, $\log{g}=3.5$, reddenened by A$_\mathrm{V}\sim1$\,mag provides a good match and minimizes the $\chi^2$. In general, the model spectra for 1800-2000\,K fit well (Figure \ref{fig:photosphere_fits}, bottom), with the exception of the peak at $\sim4\,\mu m$. The comparison also reveals the presence of significant excess emission above the photosphere for wavelengths beyond 4$\,\mu m$.

With an apparent J-band magnitude of 17.61, extinction of $A_V=1$, and a distance of 190\,pc the absolute magnitude is $M_J=10.97$. Comparing the temperature and absolute magnitude to the ATMO2020 model isochrones \citep{Phillips2020NewSetAtmosphere} with ages of 1-5\,Myr yields a mass between 6 and 10 Jupiter masses, again in agreement with previous estimates. 

\subsection{Accretion rate}

\begin{figure*}[ht!]
\centering
\includegraphics[width=0.9\textwidth]{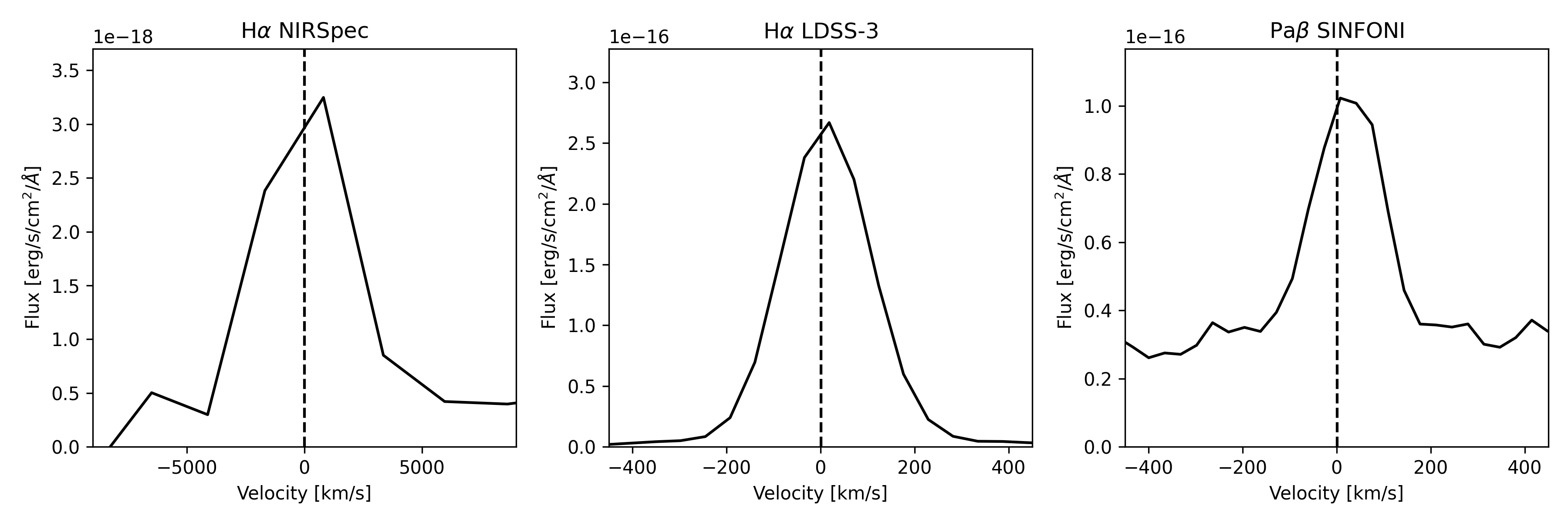}
\caption{Zoom to the region around the H$\alpha$ and Pa$\beta$ emission lines in our NIRSpec data, as well as the LDSS-3 and SINFONI spectra of \chaobj, respectively.  \added{Note that in the NIRSpec data, the width of the line is due to the low-resolution of the instrument.} Originally the spectra in the middle and right panel were published in \citet{Luhman2008DiskPopulationChamaeleon} and \citep{almendros-abadYouthAnalysisInfrared2022}. \label{fig:accretion}}
\end{figure*}

The available ground-based optical \citep{Luhman2008DiskPopulationChamaeleon} and infrared \citep{almendros-abadYouthAnalysisInfrared2022} spectra show strong H$\alpha$ and Pa$\beta$ emission, both clear signs of accretion \citep{Manara2017ExtensiveVLTXshooter}. The H$\alpha$ emission has a 10\% full width of 374 km/s, well over the typical 200 km/s accretion cutoff in BDs \citep{Jayawardhana2003EvidenceTauriPhase}. The H$\alpha$ line is also visible in our new NIRSpec data, albeit at much lower resolution. Here we derive the mass accretion rates from these lines. We flux calibrated the optical spectrum from \citet{Luhman2008DiskPopulationChamaeleon} by comparing it to the NIRSpec spectrum for $\lambda<$850 nm, avoiding regions of strong telluric absorption and the H$\alpha$ line itself. The available SINFONI spectrum is flux calibrated by comparison with JHK photometry \citep{almendros-abadYouthAnalysisInfrared2022}. We show the Hydrogen emission lines in Figure \ref{fig:accretion}.

We estimate the H$\alpha$ and Pa$\beta$ line fluxes from the published spectra with the same method used in \citet{Almendros-Abad2024EvolutionRelationMass}. We converted to line luminosities using a distance of 190\,pc. To derive accretion luminosities, we used the \citet{Alcala2017XshooterSpectroscopyYoung} relationships. In these steps all uncertainties are propagated through. We derived accretion luminosities of $\log{L_\mathrm{acc}}=-3.8\pm0.6 \,L_\odot$ from Pa$\beta$ and $-4.6\pm0.3\,L_\odot$ from H$\alpha$. Lastly, we estimated the mass accretion rate assuming a truncation radius of 5$R_*$ and an object radius of 0.21 $R_\odot$, using $\dot{M}_\mathrm{acc}=1.25\cdot L_\mathrm{acc}R_*/(GM_*)$. We find $\log{\dot{M}_\mathrm{acc}}=-9.7\pm0.8\,M_\odot$/yr using Pa$\beta$, and $-10.6\pm0.4\,M_\odot$/yr using H$\alpha$. The Pa$\beta$ accretion estimate is larger than using H$\alpha$, but the two values are consistent with each other within the errors. 

As far as we are aware, \chaobj\ is the lowest mass isolated object with confirmed gas accretion observed in optical and near-infrared emission lines. 
Measurements of accretion rates in sources of similar mass (SpT$\geq$M9) are typically $\leq10^{-10} - 10^{-10.5}\,M_\odot /yr$ \citep{Mohanty2005TauriPhaseNearly, HerczegUVExcessMeasures2008, Joergens2013OTS44Disk, Petrus2020NewTakeLowmass, Almendros-Abad2024EvolutionRelationMass}. Similarly, \citet{Viswanath2024ExoplaNeTAccRetionMOnitoring} report an accretion rate of $1.4 \cdot 10^{-11}\,M_\odot$/yr for a young L2 brown dwarf with a mass straddling the Deuterium-burning limit. 

\begin{figure}[ht!]
\centering
\includegraphics[width=0.5\textwidth]{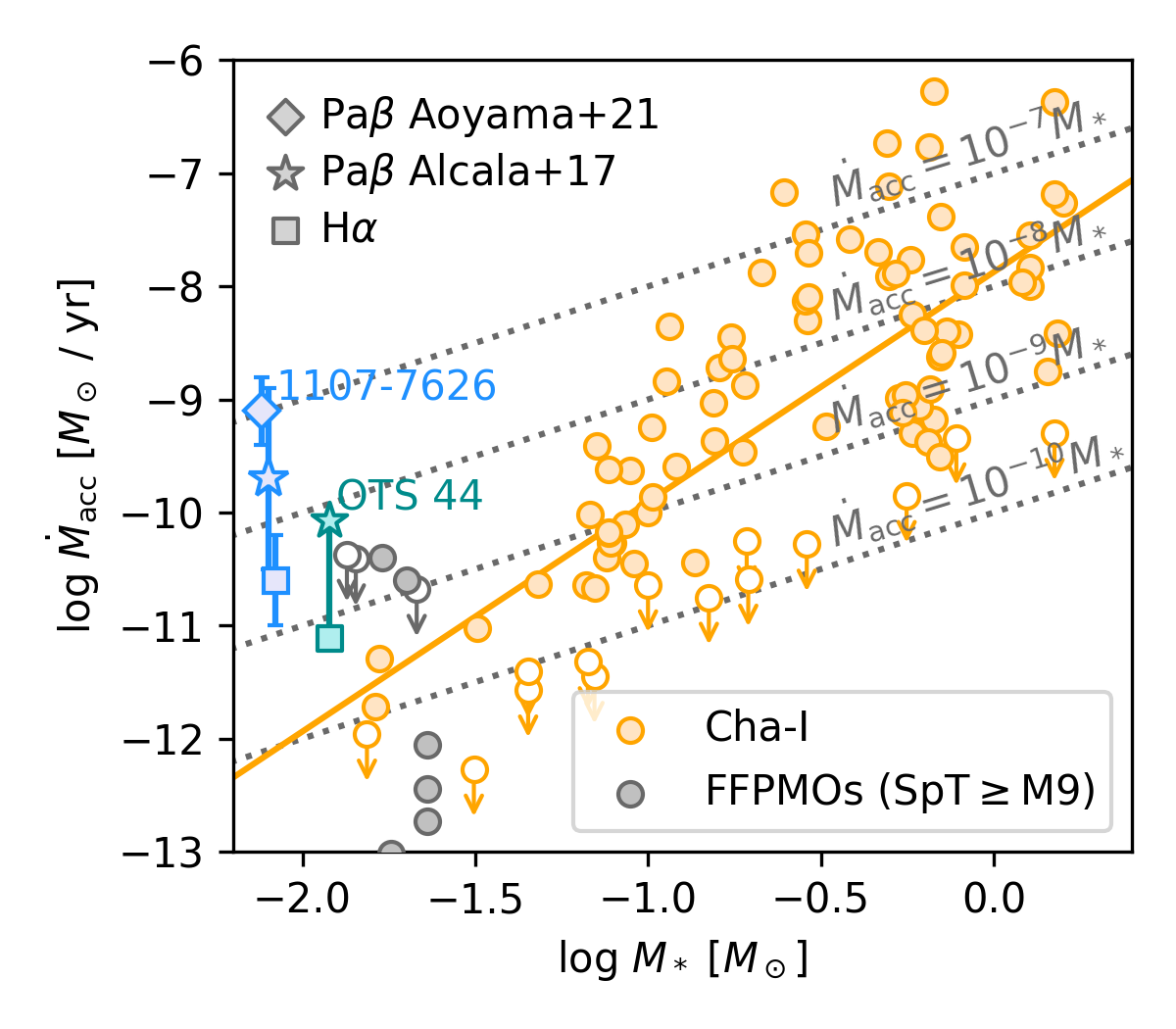}
\caption{\added{Relationship between $M_*$ and $\dot{M}_\mathrm{acc}$ for \chaobj\ (blue symbols, this work), OTS 44 \citep[cyan symbols,][]{Joergens2013OTS44Disk}, remaining Cha-I members with accretion estimate available \citep[orange circles,][]{manaraDemographicsYoungStars2023, Almendros-Abad2024EvolutionRelationMass}, and other FFPMOs \citep{BettiComprehensiveArchiveSubstellar2023, Almendros-Abad2024EvolutionRelationMass}. The solid orange line represents the correlation between these parameters found for Cha-I members in \citet{Almendros-Abad2024EvolutionRelationMass}.  \chaobj\  has an unusually high accretion rate given its mass in comparison with other low mass stars and brown dwarfs.  Additionally, the rate measured from H$\alpha$ is higher than that from  measure Pa$\beta$.}\label{fig:accretion_comp}}
\end{figure}

Compared to these values, our target is quite strongly accreting for its mass, and this is especially significant in the case of the measurement derived from Pa$\beta$ (Figure \ref{fig:accretion_comp}). A similar case may be OTS 44, a M9.5 member of Cha-I. \citet{Joergens2013OTS44Disk} estimated accretion rates of OTS 44 and find that the value from Pa$\beta$ is about two orders of magnitudes larger than the one from H$\alpha$. They concluded that Pa$\beta$ emission may have an origin different from accretion, but do not specify an alternative. The measurements OTS44 and \chaobj\ may indicate that the empirical methodolodogy to determine accretion rates may need to be reconsidered for the planetary-mass domain.

Recently, \citet{Hashimoto2024AnalysesMultipleBalmer} have found that the hydrogen lines of a non-negligible fraction of free-floating substellar objects are better reproduced by ``accretion shock" models, rather than ``accretion flow" models developed for accreting stars. \citet{AoyamaComparisonPlanetaryHaemission2021} developed relationships between line and accretion luminosities in the context of an ``accretion-shock" model. Using these relations, we estimate a mass accretion rate $\log{\dot{M}_\mathrm{acc}}=-9.1\pm0.3\,M_\odot$/yr using Pa$\beta$, and $-9.7\pm0.3\,M_\odot$/yr using H$\alpha$. Thus, these models predict an accretion rate 0.6-0.9 dex larger than from the conventional procedure and would make it even more discrepant from the trends found in the stellar regime.

\section{The disk emission}
\label{sec:disk}

\subsection{Overview}

In Figure \ref{fig:spec}, upper panel, we show our MIRI mid-infrared spectrum for \chaobj\ (black) in comparison with a photospheric model (bright blue, see Section \ref{sec:para}). As can be appreciated from this figure, the object shows substantial excess emission in the infrared indicating the presence of a disk. At wavelengths $>8\,\mu m$, the emission from the disk dominates over the photosphere.

The MIRI spectrum also shows two notable spectral features at approximately 7.7 and 10.5$\,\mu m$ (highlighted in Figure \ref{fig:phot}). Comparing with the literature on molecular emission in disks around young stars, these features can be clearly identified as methane (\methane) and ethylene (\ethylene) emission from the disk \citep{arabhaviAbundantHydrocarbonsDisk2024}. There is a hint of emission at 12-12.5 \microns\ that could be due to C$_2$H$_6$ (ethane), but the low signal-to-noise prevents a definitive detection at this time.  We see no evidence of CO, H$_2$O, or C$_2$H$_2$.

In Figure \ref{fig:spec} we also show the MIRI/MRS spectrum of ISO-ChaI 147 published by \citet{arabhaviAbundantHydrocarbonsDisk2024}, scaled by a constant factor chosen to match the flux levels in our spectrum for \chaobj.  ISO-ChaI 147 also shows the same two emission features in its mid-infrared spectrum. ISO-ChaI 147 is  a classical T Tauri star with a mass of 0.11 M$_\sun$ \citep{PascucciSteeperLinearDisk2016}, about 10-20 times more massive than our target, and about two orders of magnitude more luminous. Despite these differences, the spectra match remarkably well, both in the continuum and in the emission features. This illustrates that structure and chemical evolution of the inner disks can be self-similar, across a wide range of central object masses. 

\subsection{Modeling the gas disk emission}

\begin{figure}[ht!]
\centering
\includegraphics[width=0.5\textwidth]{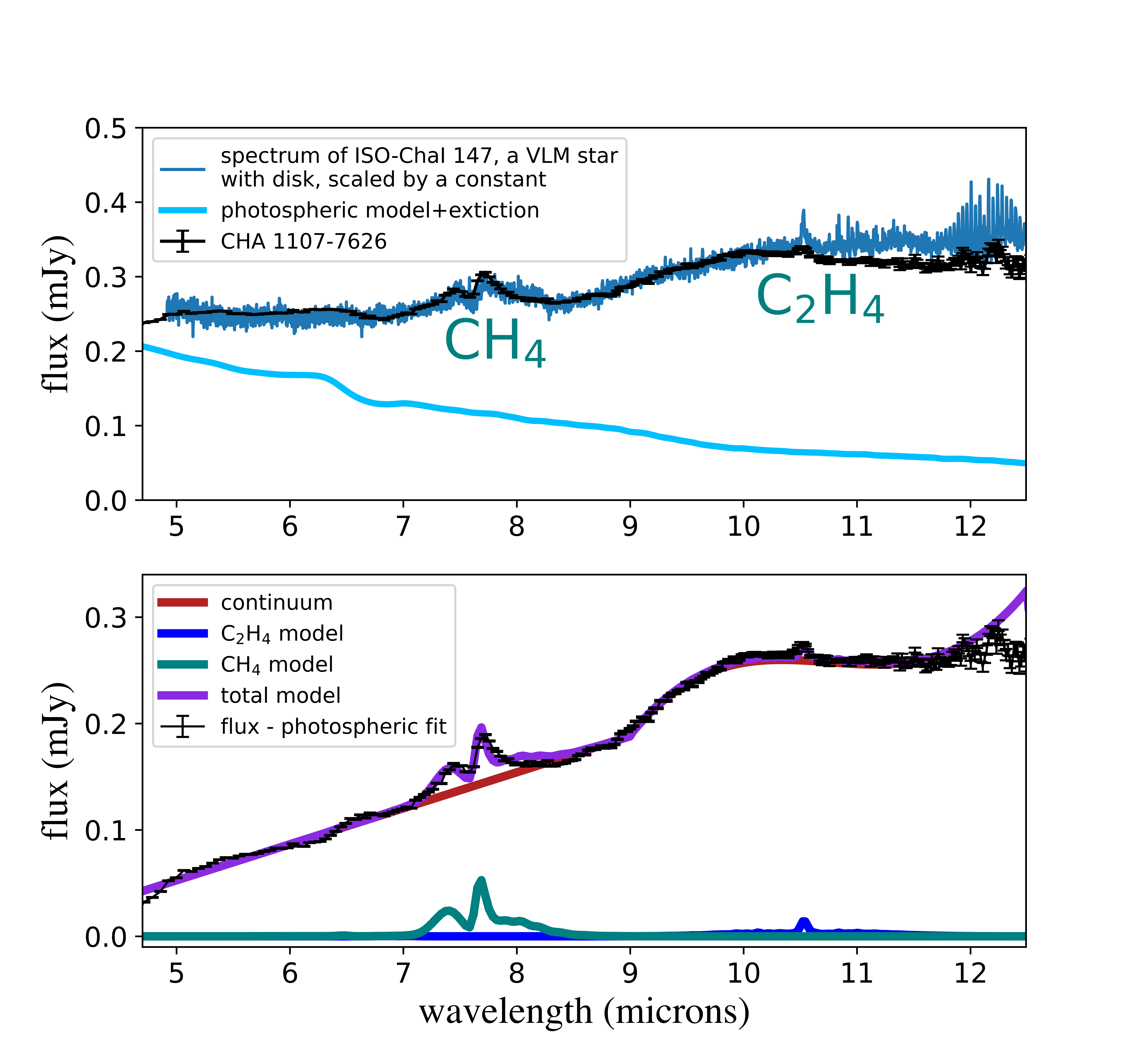}
\caption{{\bf Top panel:} The JWST spectrum of \chaobj (black), its best-fit  photospheric model (light blue), corrected for extinction. In blue we overplot the MIRI-MRS spectrum of ISO-ChaI 147, scaled by a constant. {\bf Bottom panel:} The MIRI spectrum minus the best-fit photosphere (black), and the best fit models for the continuum (dark red), \methane\ (green), and  \ethylene\ (blue).  Also plotted is the sum of those models (purple).  \added{Data from this figure is available as Data behind the Figure.} \label{fig:spec}}
\end{figure}

To further analyze the molecular emission features, we used \textit{slabspec} \citep{Salyk2022CsalykSpectools_irFirst} to create LTE slab models of \methane\ and \ethylene that would reproduce the excess emission i.e. the data minus the best-fit photospheric model. To remove the disk continuum around the methane feature, we matched the spectrum from 5.5 to 7.0 \microns\ and from 8.5 to 9.0 \microns\ with a linear function. 
Around the ethylene feature we subtracted a 3rd order polynomial fit to remove the continuum, which may be contaminated by silicate emission (see the analysis in Damian et al. 2025, in prep). 
The fit for the continuum is shown in Figure \ref{fig:spec}, bottom panel, in red. 

Our model slab spectra were created with a grid ranging in column densities from 10$^{16}$ to 10$^{20}$ cm$^{-2}$, in temperatures from 100 to 1300 K, and emitting areas from 10$^{20}$ to 10$^{23}$ cm$^{2}$. We used all isotopes  with available line lists and fixed the relative abundances for those isotopes to the ratios from HITRAN \citep{Gordon2022HITRAN2020MolecularSpectroscopic}. We then convolved the models with a gaussian to match the approximate resolution of each section. For each model we calculate the reduced $\chi^2$, designated as $\chi^2_\nu$ \citep{BevingtonDataReductionError2003}. The best fitting models are obtained for a gas temperature of a few hundred Kelvin. An example of the outcomes of the comparison with the slab models is shown in Figure \ref{fig:spec}, lower panel.

Our modeling conclusively demonstrates that the features we see are indeed \methane\ and \ethylene, as shown in Figure \ref{fig:spec}. However, the match between observed features and models is poor with regard to $\chi^2_\nu$. There are several potential reasons for the mismatch. Firstly, we know for certain that the uncertainties we use under-represent the true uncertainty (see Section \ref{sec:obs}).  We also assume a single temperature slab and a simple non-physical model for the continuum flux. 
Finally, there could be other species with opacities at the relevant wavelengths that are contributing to the spectral features we see. For example, the \methane\ emission is known to be blended with a C$_2$H$_2$ feature. For these reasons, we cannot put stringent constraints on the physical properties of the gas.

\section{Discussion and summary}

In this paper we show a complete infrared spectrum from 1 to 12$\,\mu m$ for \chaobj, a young free-floating planetary-mass object in the Chamaeleon-I star forming region \citep{Luhman2008DiskPopulationChamaeleon}. Using our new data, we rederive the properties of the object, and estimate a temperature of $\sim 1900$\,K and a mass of 6-10 Jupiter masses. In previously published spectra, as well as our JWST data, the object shows clear Hydrogen recombination emission lines indicating ongoing accretion. We estimate an accretion rate of $10^{-10}$ to $10^{-11}$ solar masses per year, comparable to higher mass brown dwarfs. This is the lowest mass isolated object with confirmed accretion in multiple lines in the optical and infrared.  The accretion luminosity is $\sim$8\% of the bolometric luminosity.

In the mid-infrared, \chaobj\ exhibits two obvious emission line features at 7.7 and 10.5$\,\mu m$, which we attribute to emission from methane and ethylene, respectively. There is hint of possible ethane emission at 12-12.5$\,\mu m$. Based on a comparison with slab models, we find that the temperature of the molecular gas that is causing these lines is likely to be in the range of a few hundred Kelvin. With only two lines, these constraints are not very restrictive, but they are reasonable in comparison with more extensive modeling for hydrocarbon lines seen in disks around more massive stars.  

The emission line features present in the infrared spectrum of \chaobj\ presents us with an opportunity to study gas accretion and the physical conditions in the warm inner gaseous disk in depth, for the first time for an object with a mass below 10 times the mass of Jupiter. This makes \chaobj\ a compelling target for follow-up observations.

The mid-infrared spectra of disks around young stars display great variety, implying substantial diversity in the underlying physical conditions. The presence of hydrocarbons is indicative of a carbon-rich chemistry, and a high C/O ratio in the gas phase \citep{kanwarMINDSHydrocarbonsDetected2024}. The hydrocarbon features discussed seem to be common in disks around very low mass stars \citep{Pascucci2013AtomicMolecularContent, arabhaviAbundantHydrocarbonsDisk2024,taboneRichHydrocarbonChemistry2023,Arabhavi2025MINDSVeryLowmass}, including one with an age of 30\,Myr \citep{Long2025FirstJWSTView}. \added{Similarly, the lack of oxygen-bearing gas species is indicative of a high C/O ratio and a low oxygen abundance in the gaseous component of the disk, similar to what is seen in disks around very low-mass stars \citep[e.g.][]{ arabhaviAbundantHydrocarbonsDisk2024,taboneRichHydrocarbonChemistry2023}.}

The mid-infrared spectrum for \chaobj\ looks remarkably similar to that of ISO-ChaI 147, a very low mass star in the same star forming region (Figure \ref{fig:spec}). Both the continuum and the emission lines in the spectrum of \chaobj\ are well reproduced by simply scaling the spectrum for ISO-ChaI 147 by a constant factor. These comparisons indicates that disk evolution processes and disk chemistry are similar across a wide range of stellar/substellar masses and luminosities. 

Numerous studies focused on the dust component have stressed that brown dwarf disks evolve following an overall phenomenology that is comparable to disks around stars \citep{Luhman2012FormationEarlyEvolution}. With our new findings, we begin to see that the same may apply to the gas in the inner parts of the disks. Disks around planetary-mass objects, with masses less than 1\% the mass of the Sun, can harbor the same array of molecular line emission that JWST observations have recently revealed in disks around low-mass stars. In particular, our object shows clear signs of a carbon-rich chemistry. This is a further indication that the evolutionary processes in disks are robust across several orders of magnitude in mass. The disk of \chaobj\ may present us with clues that isolated planetary-mass objects can form their own retinues -- raising the prospect of miniature planetary systems in their midst.

\section*{Acknowledgements}

We would like to thank the anonymous referee for their helpful comment.s  We thank Kevin Luhman who made his optical spectrum available to us. LF and RJ acknowledge support for the JWST-GO-04583.008 program provided by NASA through a grant from the Space Telescope Science Institute, which is operated by the Association of Universities for Research in Astronomy, Inc., under NASA contract NAS 5-03127.
AS and BD acknowledge support from the UKRI Science and Technology Facilities Council through grant ST/Y001419/1/. KM acknowledges support from the Fundação para a Ciência e a Tecnologia (FCT) through the CEEC-individual contract 2022.03809.CEECIND and research grants UIDB/04434/2020 and
UIDP/04434/2020. 

\bibliography{MyLibrary}{}
\bibliographystyle{aasjournal}



\end{document}